\documentclass[a4paper]{jpconf}
\usepackage{graphicx}
\usepackage[intlimits]{amsmath}
\usepackage{amssymb}
\usepackage{graphicx}
\usepackage{booktabs}
\usepackage{mathrsfs,slashed}
\usepackage{hyperref}
\usepackage{epsfig}
\usepackage{color}

\newcommand{\bea}{\begin{eqnarray}}
\newcommand{\eea}{\end{eqnarray}}
\newcommand{\beq}{\begin{equation}}
\newcommand{\eeq}{\end{equation}}

\newcommand{\nn}{\nonumber}

\newcommand{\ii}{{\rm i}}

\DeclareMathOperator{\IM}{Im}

\begin{document}
\title{ Pion and kaon in the Beth-Uhlenbeck approach}

\author{A.~Dubinin${}^{1}$,~D.~Blaschke${}^{1,2,3}$,~A.~Radzhabov${}^{4}$ }

\address{${}^{1}$ Instytut Fizyki Teoretycznej, Uniwersytet
Wroc{\l}awski, 50-204 Wroc{\l}aw, Poland}
\address{${}^{2}$ Bogoliubov Laboratory for Theoretical Physics, JINR Dubna, 141980 Dubna, Russia}
\address{${}^{3}$ National Research Nuclear University (MEPhI), 115409 Moscow, Russia  }
\address{${}^{4}$ Matrosov Institute for System Dynamics and Control Theory, Irkutsk 664033, Russia}

\ead{aleksandr.dubinin@ift.uni.wroc.pl}

\begin{abstract}
In the present work the Mott effect for pions and kaons is described within a Beth-Uhlenbeck approach on the basis of the PNJL model. The contribution of these degrees of freedom to the thermodynamics is encoded in the temperature dependence of their phase shifts. A comparison with results from $N_f=2+1$ lattice QCD thermodynamics is performed.
\end{abstract}

\section{Introduction}

One of the central problems in the investigation of the transition from  hadronic to quark matter is a microphysical description of the dissociation of hadrons into their quark constituents. This Mott transition occurs under extreme conditions of high temperatures and densities as they are provided, e.g., in  ultrarelativistic  heavy-ion collisions or in the interiors of compact stars. Since an ab-initio description of QCD thermodynamics within simulations of lattice QCD (LQCD) is yet limited to finite temperatures and low chemical potentials only, the development of effective model descriptions is of importance. Here, we develop a relativistic Beth-Uhlenbeck approach to the description of mesonic bound and scattering  states in a quark plasma \cite{Hufner:1994ma,Blaschke:2013zaa} further by including the strange sector. 
To this end we employ the PNJL model which is particularly suitable for addressing the appearance of pions and kaons as both, quasi Goldstone bosons of the broken chiral symmetry and pseudoscalar meson bound states. 
Within this framework the confinement  of colored quark states is effectively taken into account by coupling the chiral quark dynamics to the Polyakov loop and its effective potential. The model is widely used to describe quark-gluon thermodynamics in the meanfield approximation \cite{Fukushima:2003fw,Ratti:2005jh}, but has also been developed to address mesonic correlations 
\cite{Hansen:2006ee,Blaschke:2007np,Radzhabov:2010dd,Wergieluk:2012gd,Yamazaki:2012ux,Yamazaki:2013yua,Blaschke:2014zsa,Blaschke:2015nma}. 
The relativistic Beth-Uhlenbeck approach is the appropriate tool to develop a unified description of quark-gluon and hadron thermodynamics including the transition between both asymptotic regimes of QCD.

In the next section the basic formulae for the thermodynamic potential, the phase shifts and the pressure in the Beth-Uhlenbeck approach  are presented, followed by a discussion of numerical results compared to LQCD data in Sect.~3 and conclusions in Sect.~4.

\section{Relativistic Beth-Uhlenbeck approach for the PNJL model}
The Lagrangian for the 3 flavor PNJL model is given by
\bea
\mathscr{L}
	&=&
	\bar{q}[\ii\slashed\partial - m_0+\gamma_0(\mu-\ii A_4)]q+G_{\rm S}\sum^{8}_{a=0}\left[(\bar{q}\lambda^{a} q)^2+(\bar{q}\ii\gamma_5\lambda^{a} q)^2\right]
	-\mathscr{U}(\Phi, \Bar{\Phi}; T)~.\nn
\eea
Here, $q$ denote the  quark fields with 3 colors and flavors,  $\lambda^a$ are  the  Gell-Mann matrices in flavor space  $(a=0,1,2,3\ldots8)$, $G_S$ is dimensionful coupling constant. The Polyakov-loop potential $\mathscr{U}(\Phi, \Bar{\Phi}; T)$ {\color{black} is chosen} in the polynomial form  \cite{Ratti:2005jh}, with the parameters taken from that reference.
The gluon background field in the Polyakov gauge is a diagonal matrix in color space
$A_4 = {\rm diag}(\phi_3+\phi_8,-\phi_3+\phi_8,-2\phi_8)$. 
The Polyakov loop field $\Phi$ is defined via the color trace over the gauge-invariant average of the Polyakov line $L(\vec{x})$ \cite{Ratti:2005jh}.

The thermodynamic potential in Gaussian approximation has the form \cite{Blaschke:2014zsa}  
\bea
\Omega_{\rm Gau{\ss}}=	\mathscr{U}(\Phi, \Bar{\Phi}; T)
	+\frac{\sigma_{\rm MF}^2}{4G_{\rm S}}
	+	\Omega_{\rm Q}~
	+   \Omega_{\rm M}~,
\eea

where in the quark $(\Omega_{\rm Q})$ and meson $(\Omega_{\rm M})$ contributions the zero-point energy terms are removed ("no sea" approximation). 
The quark contribution is given by
\bea
\label{Omega_Q-f}
\Omega_{\rm Q}
	&=&
-\frac{2N_c N_f}{3}\int\frac{dp}{2\pi^2}\frac{p^4}{E_p}
	\left[f_\Phi^+(E_p) + f_\Phi^-(E_p)\right]~,
\eea
where $f_\Phi^{\pm}(E_p)$ are the generalized Fermi distribution functions
\bea
\label{f_Phi}
f^+_\Phi(E_p)=
\frac{(\bar{\Phi}+2{\Phi}Y)Y+Y^3}{1+3(\bar{\Phi}+{\Phi}Y)Y+Y^3}
~,~
f^-_\Phi(E_p)=
\frac{({\Phi}+2\bar{\Phi}\bar{Y})\bar{Y}+\bar{Y}^3}{1+3({\Phi}+\bar{\Phi}\bar{Y})\bar{Y}+\bar{Y}^3}~.
\eea
with the abbreviation $Y={\rm e}^{-(E_p-\mu)/T}$ and $\bar{Y}={\rm e}^{-(E_p+\mu)/T}$
(cf. Ref.~\cite{Hansen:2006ee}).
The meson contribution reads \cite{Blaschke:2013zaa}
\bea
\label{Omega_D-g}
\Omega_{\rm M} &=&
-d_{M} \int\frac{d^3q}{(2\pi)^3}\hspace{-2mm}\int\frac{d\omega}{2\pi}
	\left[g^+(\omega+\mu_M) + g^-(\omega-
	\mu_M)\right] \delta_M(\omega,\bf q)~.
	\nonumber
\eea
Here  $d_{M}$ are the meson degeneracy factors: $d_{\pi}=3$ for pions and $d_{K}=4$ for kaons;~  $g(E)=(\e^{\beta E}-1)^{-1}$ is the Bose function, $\mu_M = \mu_i-\mu_j$ is the chemical potential of a meson M composed of quark $i$ with chemical potential $\mu_i$ and
antiquark $j$ with chemical potential $-\mu_j$. 
The phase shift at rest $\delta_{\rm M}(\omega, {\bf q=0})=\delta_{\rm M}(\omega)$ is determined by
\begin{equation}
\label{phaseshift}
\delta_{\rm M}(\omega)=
-\IM\ln\left[\beta^2S_{\rm M}^{-1}(\omega-\mu_M+\ii\eta)\right]~.
\end{equation} 

The propagator and the polarization function in the present work should be understood as to be evaluated at the shifted energy $z=\omega-\mu_M+\ii\eta$
\bea	
	S_{\rm M}^{-1}(z)
	&=&{G_{\rm S}}^{-1} - \Pi_{\rm M}(z)~,\\
\Pi_{\rm M}(z) &=& 
4\{\Big(I^{i}_{1}+I^{j}_{1}\Big)-
[ (\omega-\mu_M)^2-(M_i-M_j)^2]I^{ij}_{2}(\omega-\mu_M) ]~,\nn
\label{corr-prop}
\eea
where the 1-loop integrals are given by
\bea
I^{i}_{1}= -N_c\int\frac{dp}{4\pi^2}\frac{p^2}{E_i}\left[f_\Phi^+(E_i) + f_\Phi^-(E_i)\right]~,
\eea
\hspace{-10 mm}
\bea
I^{ij}_{2}= -N_c\int \frac{d^3p}{(2\pi)^3}
\Big[ 
\frac{1}{2E_i}
\frac{1}{(E_i+\omega)^2-E^2_J}f_\Phi^+(E_i)
- \frac{1}{2E_i}
\frac{1}{(E_i-\omega)^2-E^2_J}f_\Phi^-(E_i) \nn\\
+\frac{1}{2E_j}\frac{1}{(E_j-\omega)^2-E^2_i}f_\Phi^+(E_j)- \frac{1}{2E_j}\frac{1}{(E_j+\omega)^2-E^2_i}f_\Phi^-(E_j)\Big]~.
\eea

\section{Results}
The parameters used for the numerical studies in the present work are the bare quark masses $m_{u,d} = 5.5$~MeV and $m_{s} = 138.6$~MeV, the three-momentum cutoff $\Lambda = 602$~MeV and the scalar coupling constant $G_{\rm S}\Lambda^2 = 2.317$. With these parameters one finds in vacuum a constituent quark mass of 367~MeV, a pion mass of 135~MeV and pion decay constant $f_\pi=92.4$ MeV.
We report here results for the case of vanishing chemical potentials.

\begin{figure}[!htb]
\includegraphics[width=0.5\textwidth,height=0.4\textwidth]{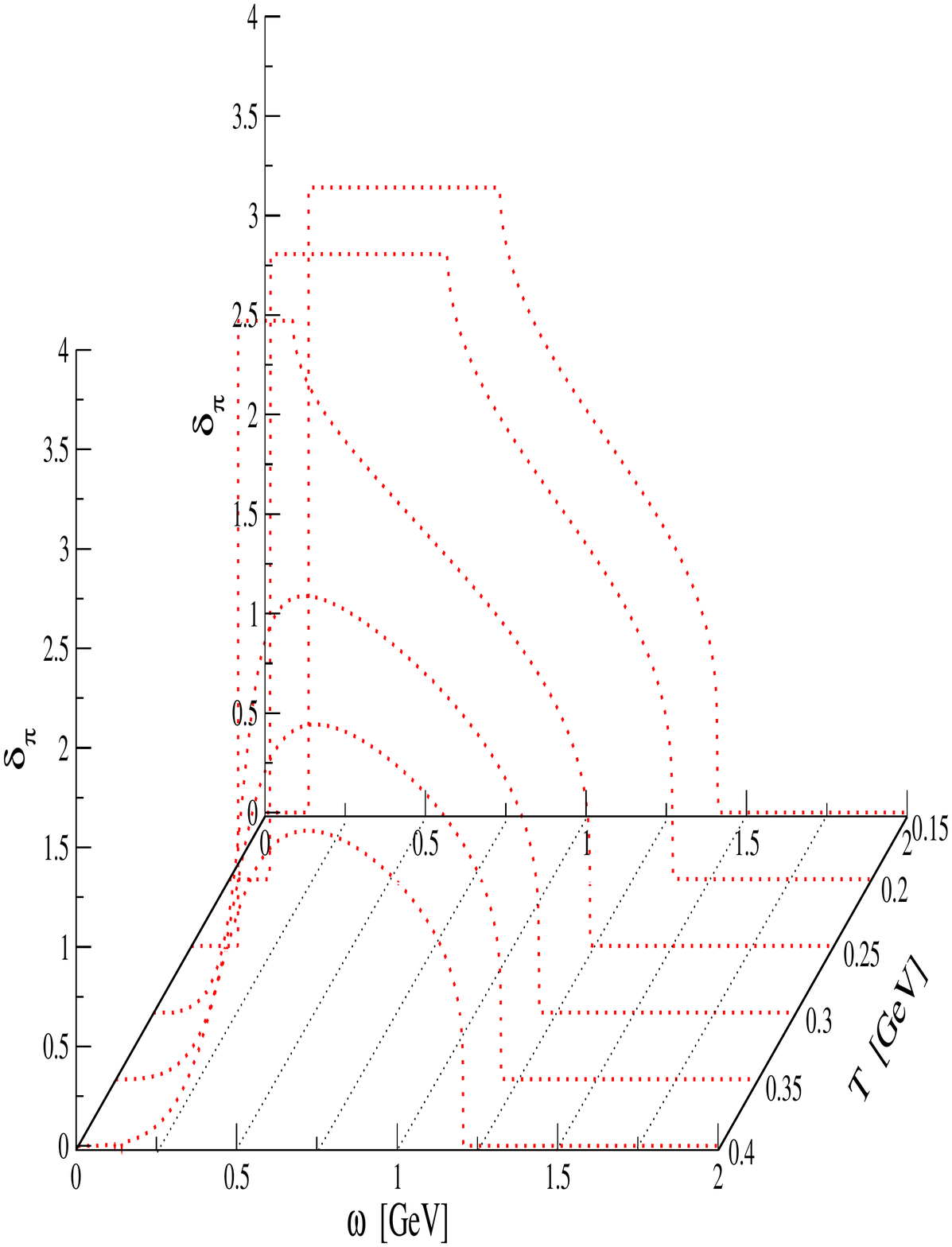}
\includegraphics[width=0.5\textwidth,height=0.4\textwidth]{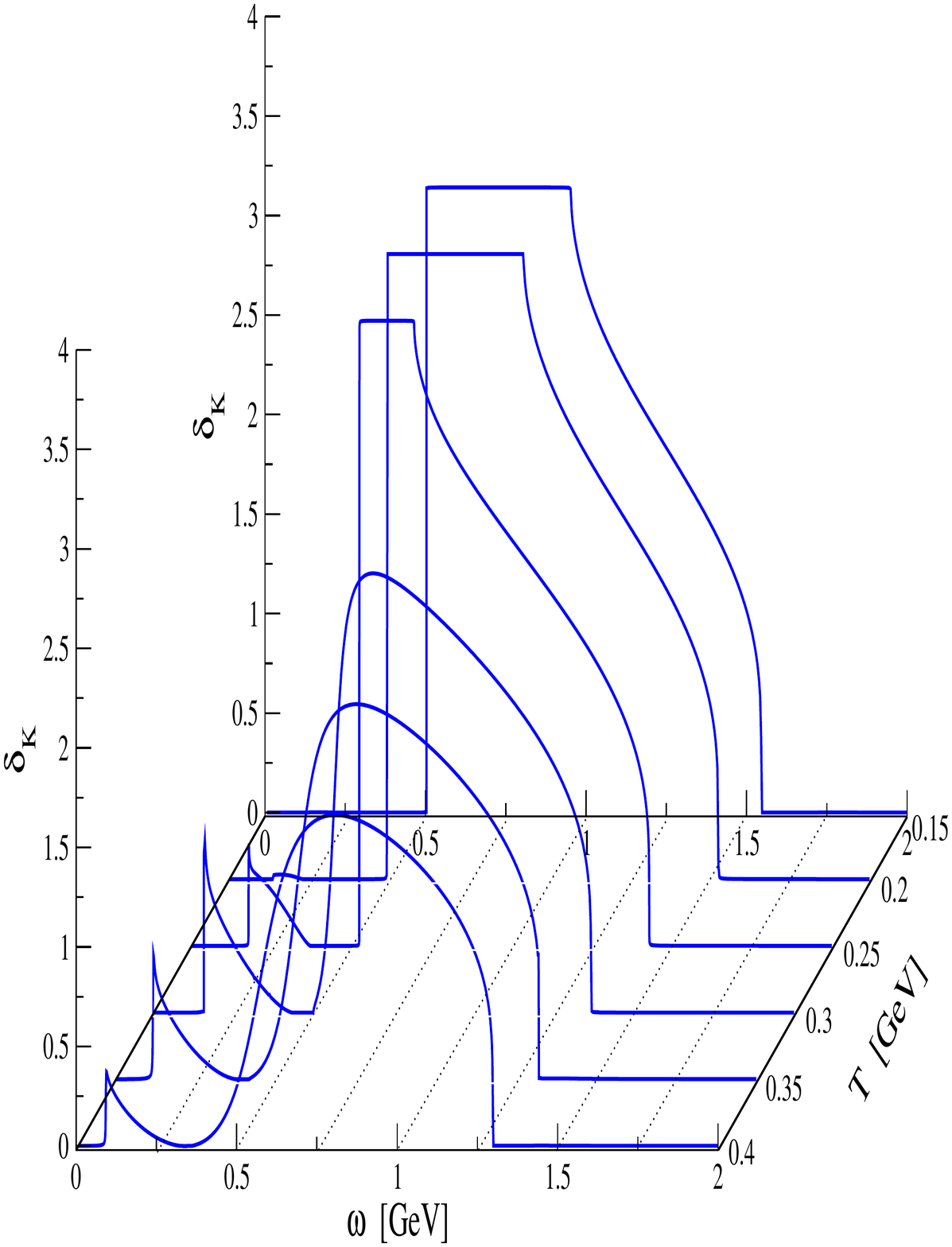}
\caption{Phase shift of pions (left panel) and kaons (right panel) as function of the energy for different temperatures from $T = 150$ MeV to $400$ MeV.}
\label{fig:ph1} 
\end{figure}

The solution for the pion phase shift is shown in  Figure~\ref{fig:ph1} as function of the energy variable 
$\omega$. 
The jump of the phase shift from $0$ to $\pi$ indicates the position of a bound state in the spectrum below the threshold of the continuum states which is located where the phase shift starts decreasing
towards zero. 
With increasing temperature the threshold moves to lower $\omega$-values and the phase shift jumps from $\pi$ to $0$ when the Mott temperature is reached where the bound state gets dissociated. 
This behavior is in accordance with the Levinson theorem, for details see \cite{Blaschke:2014zsa}.
\begin{figure}[!htb]
\includegraphics[width=0.5\textwidth]{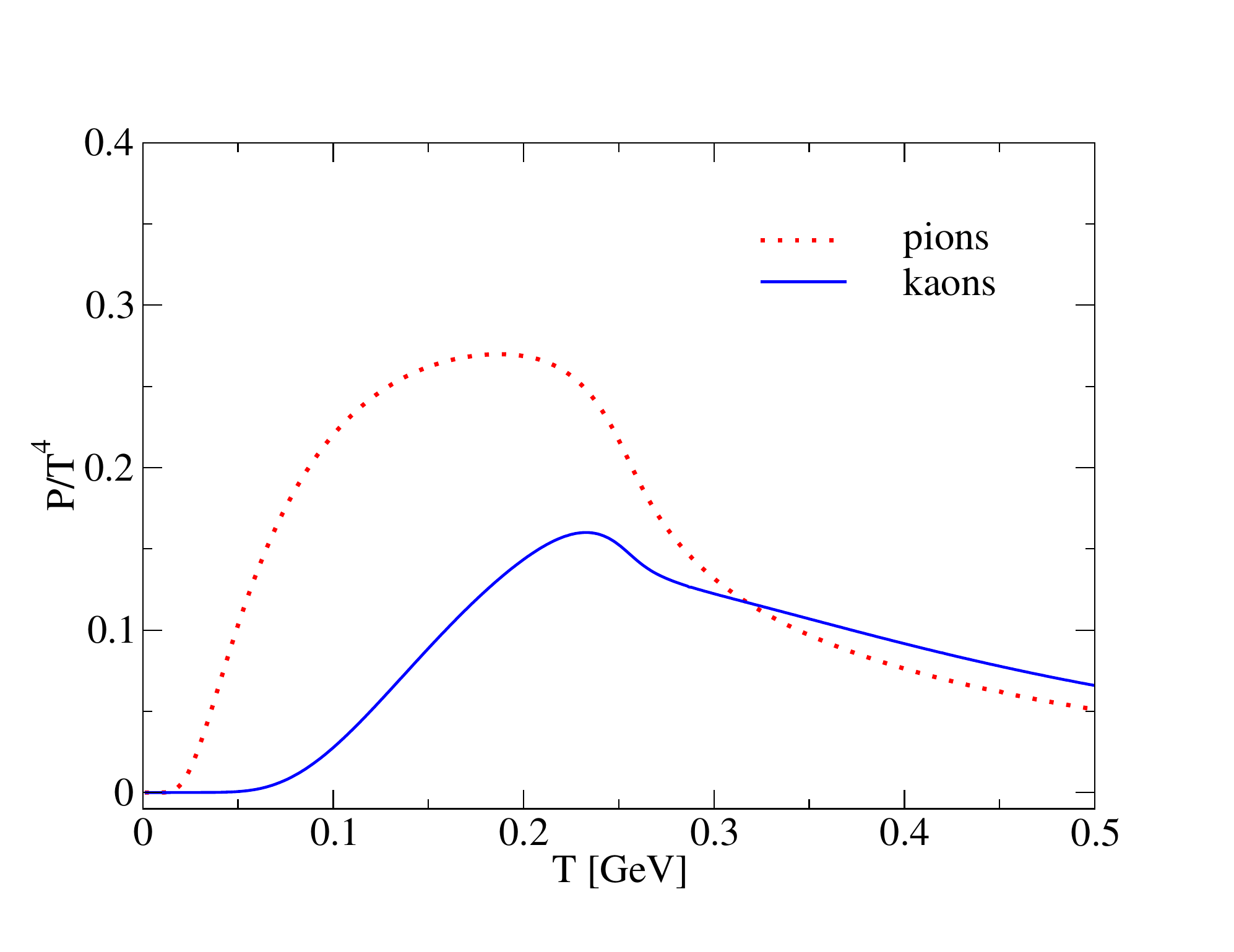}
\includegraphics[width=0.5\textwidth]{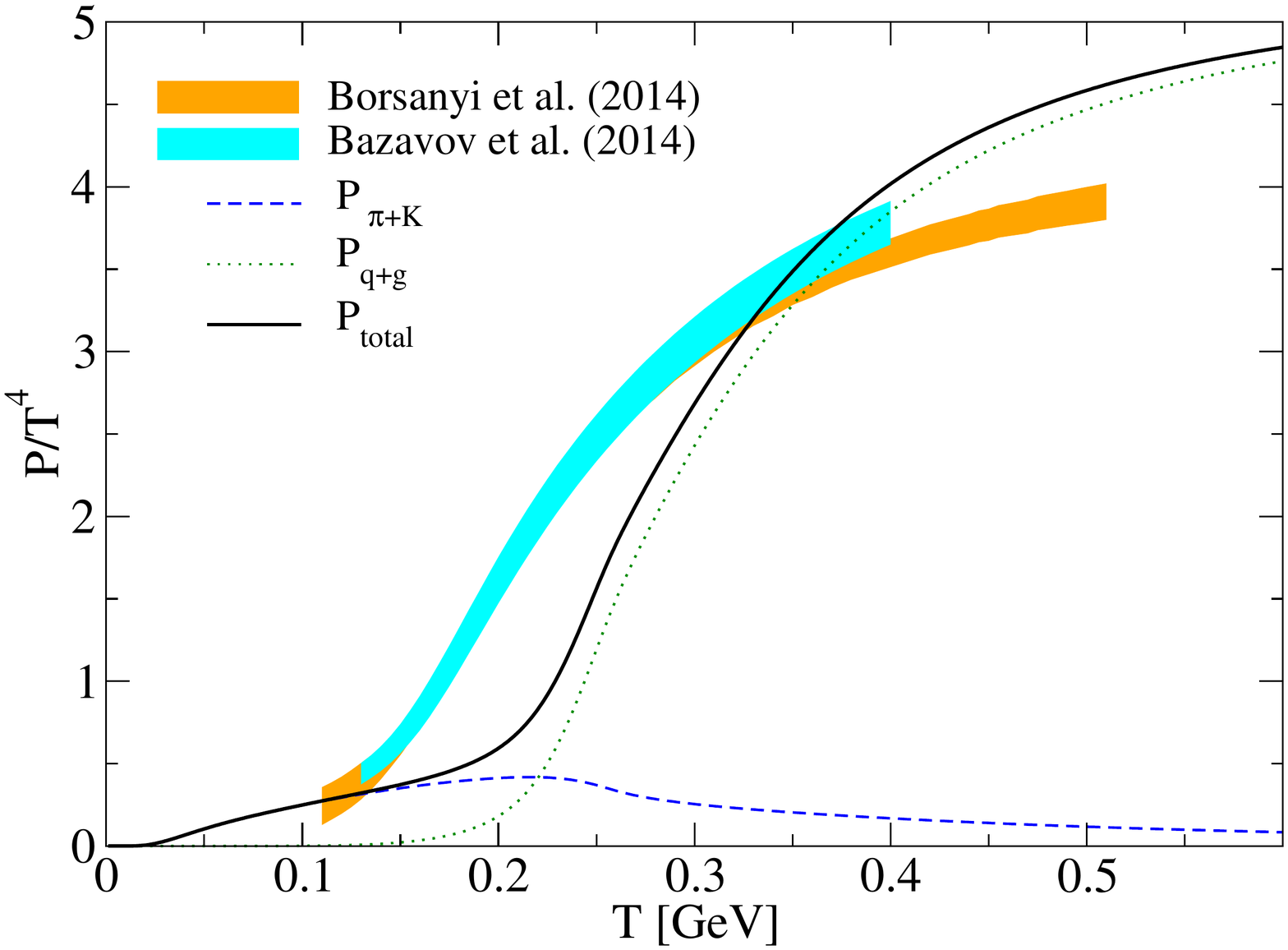}
\caption{Left panel: pressure of pions (red dotted line) and kaons (blue solid line). Right panel: total pressure of  $N_f=2+1$  flavor PNJL model with  pseudoscalar meson  correlations as a function of temperature (black solid line) and LQCD data (colored bands \cite{Borsanyi:2013bia,Bazavov:2014pvz}). }
\label{fig:pressure} 
\end{figure}
In the kaon phase shift arises one more threshold at low energy because the dynamical masses of the quarks composing the kaon are different, for more information see Ref.\cite{Yamazaki:2012ux,Yamazaki:2013yua}. 
In the left panel of Figure~\ref{fig:pressure} the pressure of pions and kaons is  presented. It  shows a typical behavior: first increasing with temperature towards the Stefan-Boltzmann limit which  isn't reached since due to the chiral phase transition the continuum threshold lowers and this induces a reduction of the meson gas pressure already before the Mott temperature is reached. 
Above the Mott temperature, the growing meson width leads to a stronger reduction of the pressure with a rather sharp onset of this effect. 
In the right panel of  Figure~\ref{fig:pressure} we compare the total pressure of the $N_f=2+1$ flavor PNJL model with recent LQCD data from \cite{Borsanyi:2013bia,Bazavov:2014pvz}. 
While our results agree with LQCD below  $T\simeq 150$ MeV, a discrepancy opens up beyond this temperature which is mainly due the fact that the chiral transition temperature in the PNJL model is too high ($\sim 250$ MeV) when compared to the pseudocritical temperature of $T_c=154 \pm 9$ MeV from LQCD
\cite{Kaczmarek:2011zz}.  
Above a temperature of $T \simeq 300$ MeV the model pressure reaches towards the Stefan-Boltzmann limit while the LQCD pressure stays below it. 
%
\section{Conclusions}
The main result this work is the addition of the strange sector to the relativistic Beth-Uhlenbeck approach which allows to compare with $N_f=2+1$ LQCD data. 
At the present level of description, there is a discrepancy between our results and the LQCD ones for temperatures $T\gtrsim 150$ MeV which can be reduced as follows.
First, by including higher lying hadronic states into the calculation of the total pressure. 
Second, by lowering the Mott temperature. 
Third, by accounting for virial corrections (perturbative) to the quark-gluon pressure corresponding to residual interactions of quarks and gluons in the plasma. 
These improvements are subject of our current work. 
%
\section*{Acknowledgement}		
We gratefully acknowledge discussions about lattice QCD data with O. Kaczmarek.
The work of A.D. was supported by the Polish National Science Centre  (NCN) under grant number UMO - 2013/09/B/ST2/01560  and by the Institute for Theoretical Physics of the University of Wroclaw under internal grant number 1439/M/IFT/15. D.B. and A.R.  received support by NCN under grant number UMO-2011/02/A/ST2/00306.


\end{document}